\documentclass[conference]{IEEEtran}
\IEEEoverridecommandlockouts

\def\BibTeX{{\rm B\kern-.05em{\sc i\kern-.025em b}\kern-.08em
    T\kern-.1667em\lower.7ex\hbox{E}\kern-.125emX}}

\usepackage{subcaption}
\usepackage{color,soul}
\usepackage{url}
\usepackage{xurl}
\usepackage{comment}
\usepackage{graphicx}
\usepackage{hyperref}
\usepackage[margin=1in]{geometry}
\usepackage{tabularx}
\usepackage[numbers]{natbib}

\usepackage{dcolumn}

\begin{document}

\newcolumntype{d}[1]{D{.}{\cdot}{#1} }

\title{Investigating the concentration of High Yield Investment Programs in the United Kingdom} 
\date{}

\author{\IEEEauthorblockN{Sharad Agarwal}
\IEEEauthorblockA{\textit{Department of Computer Science} \\
\textit{University College London}\\
London, United Kingdom \\
sharad.agarwal.20@ucl.ac.uk}
\and
\IEEEauthorblockN{Marie Vasek}
\IEEEauthorblockA{\textit{Department of Computer Science} \\
\textit{University College London}\\
London, United Kingdom \\
m.vasek@ucl.ac.uk}
}

\maketitle

\begin{abstract}
Ponzi schemes that offer absurdly high rates of return by relying on more and more people paying into the scheme have been documented since at least the mid-1800s. Ponzi schemes have shifted online in the Internet age, and some are re-branded as HYIPs or High Yield Investment Programs. This paper focuses on understanding HYIPs' continuous presence and presents various possible reasons behind their existence in today's world. A look into the countries where these schemes purport to exist, we find that 62.89\% of all collected HYIPs claim to be in the United Kingdom (UK), and a further 55.56\% are officially registered in the UK as a `limited company' with a registration number provided by the UK Companies House, a UK agency that registers companies. We investigate other factors influencing these schemes, including the HYIPs' social media platforms and payment processors. The lifetime of the HYIPs helps to understand the success/failure of the investment schemes and helps indicate the schemes that could attract more investors. Using Cox proportional regression analysis, we find that having a valid UK address significantly affects the lifetime of an HYIP.
\end{abstract}


\section{Introduction}
\label{intro}

Ponzi schemes have been running for more than a century and were termed `Ponzi schemes' in the 1920s on the name of a swindler called \emph{Charles Ponzi}. An international reply coupon (IRC) was a coupon that could be exchanged for several priority airmail postage stamps from another country.
Ponzi received a letter in the post from a company in Spain that enclosed this IRC. He realized that he could make an enormous profit by purchasing IRCs in one country and trading them for more valuable stamps in another country as a form of arbitrage. Ponzi figured that by finding a way to trade with the coupons in a large quantity, one could become rich~\cite{ponzi:smithmuseum}. He started transferring money to representatives working for him in other countries, buying the IRCs and shipping them back to the United States. Ponzi convinced a few investors to fund his start up money, assuring them a 50\% profit in 45 days or 100\% in 90 days. This marked the beginning of the scheme that carries Ponzi’s name to this day~\cite{ponzi:smithmuseum}.

Ponzi promised investors returns for what he claimed was an investment. Ponzi used these funds from the new investors to pay fake returns to previous investors~\cite{ponzi:ussec}. The scheme lasted until August of 1920 when The Boston Post began investigating Ponzi's Securities Exchange Company~\cite{smithsonian.com_1998}. As a result of the newspaper's investigation, Ponzi was arrested by federal authorities on August 12, 1920, and charged with several counts of mail fraud.

Ponzi schemes take advantage of people who invest money by luring the investors into their scheme and convincing them to pay high profits.
They generally require an initial investment and promise too high of a rate of return. Classically, Ponzi schemes focus on attracting new investors to have an inflow of money to make promised payments to earlier investors and divert some of these invested funds for personal use instead of engaging in any legitimate investment business~\cite{ponzisec}. Almost with no legitimate income, Ponzi schemes require a consistent in-flow of money to survive. When it becomes challenging to achieve more investors or enormous numbers of existing investors withdraw, these schemes tend to collapse~\cite{ussecponzi}.

The widespread use of the Internet has shifted many businesses online, including scams like Ponzi schemes, referred to as High Yield Investment Programs in the Internet vernacular or HYIPs for short. The ruses for these stay similar to classic Ponzi schemes, like funding startups, dealing in expensive goods, and arbitrage, though some add modern twists like trading in cryptocurrencies~\cite{blanton2012rise}. These are a substantial form of online financial fraud and have affected thousands of people worldwide~\cite{fbi_ic3}. Ponzi schemes cause damage to the economy, and most countries prohibit them using strict financial fraud laws. Even though payment processors like Liberty Reserve have been shut down~\cite{justice_2016}, the fraud continues.

HYIP fraudsters generally set up a website promising an absurdly high-interest rate, like $1-2\%$ per day, disclosing almost no information about the underlying investments. They may use social media or online forums for advertising their HYIP. Another marketing tool they use involves encouraging investors to use social media to share information about an HYIP with others and providing a referral bonus in return~\cite{hyip:ussec}. It can be challenging to track down the actual account holders, making it harder for fraudsters to be held accountable~\cite{SEC:socialmedia}. The rise of digital currencies has made it much easier for operators of such websites to accept payments from anyone worldwide. With cryptocurrencies, the HYIPs take advantage of their pseudo-anonymous nature as well as their increased accessibility to new investors~\cite{vasek:fc15}. There are many active HYIP websites at any given time~\cite{moore:fc12}.

The primary reason to run an HYIP efficiently is its popularity, making everything readily available. Kit developers like Goldcoders sell complete HYIP website kits, allowing buyers to set up customized HYIP websites in a few minutes easily. Cryptocurrencies and payment processors like Perfect Money\footnote{\url{https://perfectmoney.com/}} are not as tightly regulated by the government, making it easy for scammers to accept money. The cost to promote HYIPs on various discussion forums and social media is negligible. The ease and accessibility of these resources have made it uncomplicated for anyone with a meager investment to start an HYIP and continue the fraud by luring more and more investors.

Scammers lure people into investing money in such schemes by promising them very high returns, cautioning a minimal risk. Ponzi schemes depend on the new investors to pay the desired profits to the already joined investors. These schemes collapse as soon as there are no new investors as there is not enough money to pay out the returns. Understanding the lifetime of the HYIPs is vital to see which schemes are successful in luring more victims. We are interested in finding new factors that influence the lifetime of the HYIPs, continuing their existence in today's era.

\subsection{Contribution}

Our work makes the following contributions:

\begin{itemize}
    \item We collect a unique updated dataset about 450 High Yield Investment Programs (HYIPs) available at \href{https://doi.org/10.7910/DVN/BLGH0A}{Harvard Dataverse}. We identify the countries where the HYIPs claim to be based, summarized in Table~\ref{tab:HYIP_distribution_countries}. (62.89\% of these HYIPs claim to be in the United Kingdom.) 
    
    \item We collect new features of analysis like address, company registration number, contact details, social media handles, and various currencies they accept through a novel process outlined in Section~\ref{sec:meth}. We also use existing variables of analysis, such as whether the HYIP is licensed by one of the most common HYIP kit developers - Goldcoders.
    
    \item We look to explain the lifetime of these scams in Section~\ref{analysis} using survival analysis on our collected variables.
    
    \item We verify HYIPs' official registration in the countries they claim to be registered as a limited company discussed in Section~\ref{sec:meth}. (55.56\% of the total investigated HYIPs are confirmed to be registered in the United Kingdom.) We also verify some artificially created credentials in the process.
    
\end{itemize}
\section{Related Work}
\label{sec:recentwork}

Moore, Han, and Clayton, in 2012, were the first researchers to investigate HYIPs~\cite{moore:fc12}. They collected data via aggregators, or services that monitor every individual HYIP, list them on their website and promote them. Moore et al. discovered that HYIPs' daily rate of return affects their lifetime and found that offering a lower rate of return (unlike 10\% or more daily return) per day attracts more investors over time and has a greater chance to live longer.

Follow-up work investigated the infrastructure supporting HYIPs. Neisius and Clayton investigated HYIP `kits', which allow potential HYIP creators to develop a highly customizable website and other necessary infrastructure with only a few clicks~\cite{neisus:apwg}.
They also investigated the link between aggregators and HYIPs. Aggregators list HYIPs to earn money through referrals by introducing investors to an HYIP.
They unveiled selective payouts, i.e., HYIPs paying only to aggregators or selected investors, scamming naive investors. The gullible investors are unaware of reporting delayed/missing payouts on the aggregators' websites, letting the schemes run longer.

Around the same time, Drew and Moore presented a new clustering method to uncover links between HYIP websites~\cite{drew:clustering}. They combined features like HTML tags and file directory structure to link websites and indicate that they belong to the same criminal.
There were some downsides in the method since they treated Goldcoders sites with different design templates as being different.
However, their work identified other groups outside of Goldcoders and other known website templates.

Nizzoli et al. provide insights about Twitter bots sending invite links to Telegram and Discord channels and found the channels to be related to Ponzi schemes by using topic modeling techniques on Telegram and Discord messages~\cite{charting_landscape}. They found 432 Telegram channels that were involved in Ponzi schemes. The research shows how scammers have been using social media platforms to recruit more investors/victims in their scheme.
Xia et al. present insights into how the coronavirus pandemic has effected various scam ecosystems~\cite{xia2020covid}. They discovered nine such coronavirus-related HYIPs. 

There has also been a variety of work on HYIPs that relates explicitly to cryptocurrencies.
Vasek and Moore investigated Bitcoin-related HYIPs in two separate studies and found that HYIPs that accepted more than just cryptocurrencies brought in more money over a more extended period of time~\cite{vasek:fc15}. They also found that scammer-victim interaction is an essential factor affecting the lifetime of the HYIP schemes~\cite{vasek:fc19}. 
Badawi et al. investigated a specific type of HYIP, a Bitcoin generator scam, which relies on small frequent payments~\cite{badawi:ieeesp20}. There has also been work by both investigative journalists and academics alike on the giant Ponzi scam, MMM, which relies on many payers around the world~\cite{buzzfeedmmm:17,boshmaf:asiaccs20}. Toyoda et al. investigated the earliest Bitcoin-related Ponzi scheme operating on the Bitcoin forums by Trent Shavers, who was later arrested for his role in the scheme~\cite{toyoda:icdmw18}.
Bartoletti et al.~\cite{bartoletti:btcdetect} and Toyoda et al.~\cite{toyoda:pattern,toyoda:btcdetect} have worked further on identifying HYIPs directly on the Bitcoin blockchain. Others have investigated HYIPs using the smart contract platform Ethereum~\cite{bartoletti:ethereum, chen:ether, chen:www}.

Clayton, Moore and Christin looked into the various concentrations found in cybercrime activities that made the authors suggest various interventions \cite{clayton2015concentrating}. In addition to economic behaviour, the authors have identified  “whack-a-mole” or inertia effects to somewhat cause an artificial concentration around HYIP scams, as well as noting that there are simply few actors in this space. They also note that cybercriminal actors often copy others successful strategies, regardless of initial intent. They propose four steps of addressing concentrations -- we note that the crucial steps here are identifying how an intervention will work and predicting the criminals' response. We address this further in our recommendations section (Section~\ref{sec:rec}).
\section{Data Collection Methodology}
\label{sec:meth}

We aim to measure the High Yield Investment Programs by collecting data from the most popular aggregator \emph{\href{https://web.archive.org/web/20191231000251/https://hyip.com/forums/review/}{hyip.com}}\footnote{\href{https://web.archive.org/web/20191231000251/https://hyip.com/forums/review/}{https://hyip.com/}}. Here, aggregators are services that monitor every individual HYIP, list them on their website, and promote them~\cite{neisus:apwg}. We chose \href{https://web.archive.org/web/20191231000251/https://hyip.com/forums/review/}{hyip.com} due to its popularity, in line with the literature~\cite{moore:fc12}. It should be noted that aggregator platforms list HYIPs as per their preferences/policies. HYIPs collected in this paper had websites available in various languages. If the website was in a non-English default language, we used the website's inbuilt language translator or Google translate to convert it to English. Except for two HYIPs, every HYIP with a non-English default language provided multiple language options.

\begin{figure}[!t]
\centering
\includegraphics[width=\linewidth]{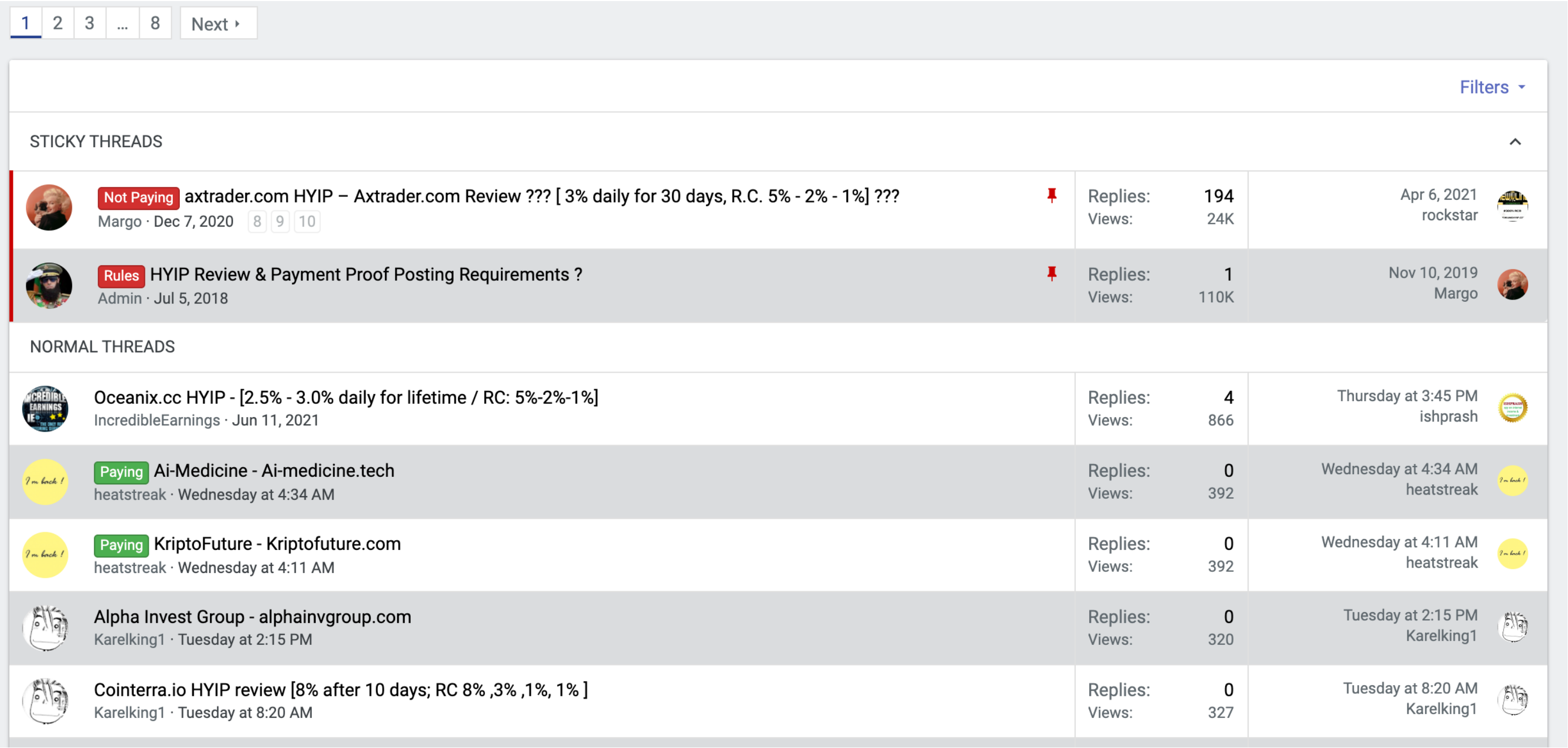}
\caption{Screenshot of the \href{https://web.archive.org/web/20191231000251/https://hyip.com/forums/review/}{HYIP Review Forum} for the aggregator \href{https://web.archive.org/web/20191231000251/https://hyip.com/forums/review/}{hyip.com}.}
\label{fig:hyipforum}
\end{figure}

\begin{figure}[!t]
\centering
\includegraphics[width=\linewidth]{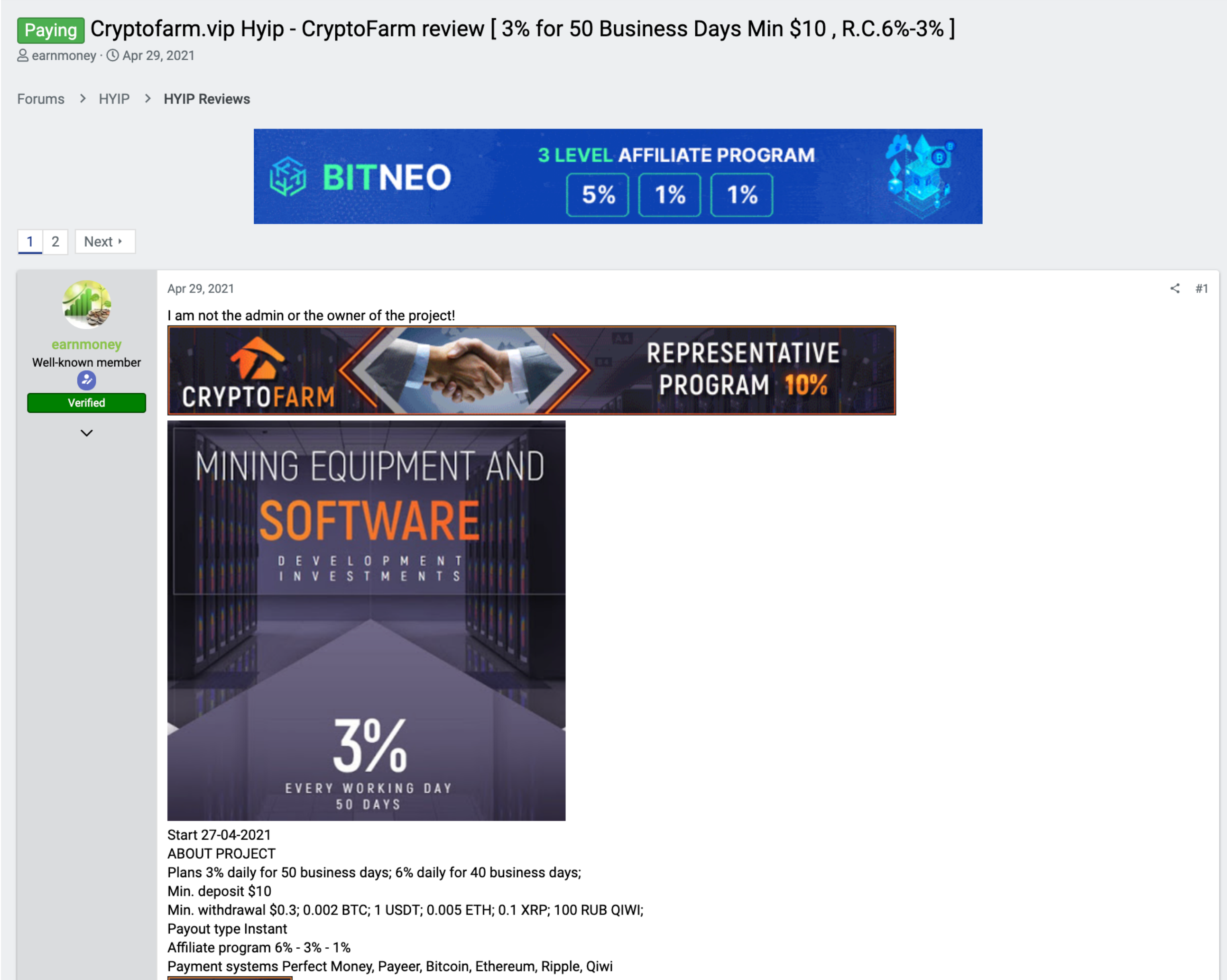}
\caption{\url{https://cryptofarm.vip/} HYIP's individual thread on HYIP Review Forum at \emph{\href{https://web.archive.org/web/20191231000251/https://hyip.com/forums/review/}{hyip.com}}.}
\label{fig:hyipindthread}
\end{figure}

The \href{https://web.archive.org/web/20191231000251/https://hyip.com/forums/review/}{hyip.com}'s website has a review forum\footnote{\href{https://web.archive.org/web/20191231000251/https://hyip.com/forums/review/}{https://hyip.com/forum/review}} that provides individual threads for every HYIP as shown in Fig~\ref{fig:hyipforum}.
The individual threads about an individual HYIP entail detailed information about the investment scheme, updates on the return on investment, and even allows investors to comment on the thread (Fig~\ref{fig:hyipindthread}).
Instead of redirecting directly to the HYIP website, it provides a link to the thread of that HYIP on another aggregator website - \emph{\href{https://web.archive.org/web/20210210124421/https://hyiprank.com/}{hyiprank.com}}\footnote{\href{https://web.archive.org/web/20210210124421/https://hyiprank.com/}{https://hyiprank.com/}}. Hyiprank provides a referral link to the HYIP website with the referrer \emph{HYIPDOTCOM}.
From this, we can assume that \emph{\href{https://web.archive.org/web/20210210124421/https://hyiprank.com/}{hyiprank.com}} is either managed by or otherwise linked in profits to \emph{\href{https://web.archive.org/web/20191231000251/https://hyip.com/forums/review/}{hyip.com}}.

Between November 2020 and September 2021, we made programmatic visits to the \emph{\href{https://web.archive.org/web/20191231000251/https://hyip.com/forums/review/}{hyip.com}} review forum and the \emph{\href{https://web.archive.org/web/20210210124421/https://hyiprank.com/}{hyiprank.com}} aggregators every alternate day (with some exceptions and minor interruptions). We parsed the data after fetching from \emph{\href{https://web.archive.org/web/20191231000251/https://hyip.com/forums/review/}{hyip.com}} review forum, to extract the URL for every individual HYIP thread on \emph{\href{https://web.archive.org/web/20191231000251/https://hyip.com/forums/review/}{hyip.com}} and \emph{\href{https://web.archive.org/web/20210210124421/https://hyiprank.com/}{hyiprank.com}} and crawled them. We then parsed the individual threads on \emph{\href{https://web.archive.org/web/20210210124421/https://hyiprank.com/}{hyiprank.com}} and \emph{\href{https://web.archive.org/web/20191231000251/https://hyip.com/forums/review/}{hyip.com}} to extract the HYIP website URL in order to crawl the HYIP web page. Our automated scraping tool is built using Selenium\footnote{\href{https://www.selenium.dev/selenium/docs/api/javascript/module/selenium-webdriver/firefox.html}{https://www.selenium.dev}} and chooses at random one user-agent disguising our browser and operating system. (We did not use Tor as we were able to scrape the web pages without any issue.)

On April $16, 2021$, the review forum was shifted under another URL -  \emph{\href{https://web.archive.org/web/20210812235219/https://www.moneymakergroup.com/forums/review/}{moneymakergroup.com}}\footnote{\href{https://web.archive.org/web/20210812235219/https://www.moneymakergroup.com/forums/review/}{https://www.moneymakergroup.com/forums/review/}}. Then, on April 19, 2021, just three days later, the \emph{\href{https://web.archive.org/web/20210210124421/https://hyiprank.com/}{hyiprank.com}} aggregator website stopped working, and the domain was not reachable. After 2-3 days, the domain started redirecting to \emph{\href{https://web.archive.org/web/20191231000251/https://hyip.com/forums/review/}{hyip.com}} and did this until September 5, 2021. Since then, the aggregator websites \emph{\href{https://web.archive.org/web/20191231000251/https://hyip.com/forums/review/}{hyip.com}}, \emph{\href{https://web.archive.org/web/20210812235219/https://www.moneymakergroup.com/forums/review/}{moneymakergroup.com}} and \emph{\href{https://web.archive.org/web/20210210124421/https://hyiprank.com/}{hyiprank.com}} stopped working. This supports our assumption that \emph{\href{https://web.archive.org/web/20191231000251/https://hyip.com/forums/review/}{hyip.com}} and \emph{\href{https://web.archive.org/web/20210210124421/https://hyiprank.com/}{hyiprank.com}} were being managed by the same entity.

\subsection{Identified Variables for Analysis} 
\label{sec:vars}
We use a variety of variables in our analysis, as we will explain below. Our source data overwhelmingly comes from the individual HYIP websites, though some also comes from the HYIP aggregator website. Table~\ref{tab:variables} summarizes our collected data.

\begin{table*}[!t]
    \centering
    \begin{tabular}{lcccrr}
    \hline
    Variable Name & Obs. & lo & hi & \multicolumn{1}{c}{Mean} & \multicolumn{1}{c}{Std Dev}\\
    \hline
    UK Address & 450 & 0 & 1 & 0.63 & 0.48\\
    Valid UK Address & 283 & 0 & 1 & 0.75 & 0.43\\
    Registered in UK & 283 & 0 & 1 & 0.88 & 0.32\\
    Same UK Address & 283 & 0 & 1 & 0.15 & 0.36\\ 

    Valid Goldcoders license & 450 & 0 & 1 & 0.64 & 0.48\\
    Payment Processors & 450 & 1 & 14 & 5.48 & 2.54\\
    Social Media Platforms & 450 & 0 & 5 & 1.20 & 1.35\\
    Contain Contact Number & 450 & 0 & 1 & 0.21 & 0.41\\
    Lifetime (days) & 450 & 0 & 1\,002 & 117.83 & 164.43\\
    \hline
\end{tabular}
\caption{\label{tab:variables}Summary of variables collected for HYIPs.}
\end{table*}

\paragraph{UK Address}
Most HYIPs provide an address on their web page. Table~\ref{tab:HYIP_distribution_countries} shows the distribution of HYIPs among various countries based on these addresses. These addresses are overwhelmingly in the UK.\footnote{We also perused two other HYIP aggregator websites - \href{https://invest-tracing.com/}{invest-tracing.com} and \href{https://www.hyip.biz/}{hyip.biz} and using a convenience sample of the most recent 15 schemes on each website, found that most HYIPs listed on these were overwhelmingly allegedly from the UK (over 73\% and 80\% HYIPs have UK addresses respectively). Because of that, we believe that our observations here are representative.} This variable is a boolean, which is true if the HYIP provides a UK address on their webpage.

\paragraph{Valid UK Address}
Most HYIPs that claim to be in the UK provides an address on their website. However, some HYIP websites do not provide an address but only the UK Companies House registration number, so we collected their addresses provided to Companies House. We used \emph{\href{https://getaddress.io/}{getaddress.io}} API to validate the addresses' existence which contains up-to-date UK addresses for all postcodes. If the HYIP has an address that is not present in the list of addresses in that postcode as per \emph{\href{https://getaddress.io/}{getaddress.io}}, we treat it as an invalid address.
This variable is true if the found address is a valid one, regardless of the further intent of the underlying address.

\paragraph{Registered in UK}
We find that 250 (55.56\%) of all analyzed HYIPs are found to be registered as a company in the United Kingdom (UK). Therefore, we mainly focus on the UK Companies House, the official government body for registering a limited company in the UK. The procedure to start a limited company in the UK is very simple. Trusting the documents provided by the individuals who want to start a company, Companies House, lacks statutory power or capability to verify the data provided while registering a company~\cite{companies_house}. The only checks conducted are to ensure the documents are complete and signed. This provides an opportunity for a scammer to register an HYIP as a limited company with the UK Companies House and attain an official registration document which they use to lure victims into investing in their schemes.

The Financial Conduct Authority (FCA) is responsible for verifying and conducting all regulated activities in the UK, including all finance management companies. The issue arises as the HYIPs register their company with the UK Companies House as a `fund management company,' `financial investment' or even as a `bank,' which is enough to make a naive individual in believing that the website is a trustworthy investment scheme and not an HYIP scam. The absence of information sharing between the FCA and the Companies House might be a reason enabling such schemes to get themselves registered. Additionally, the Companies House provides any nature of business - Standard Industrial Classification (SIC)\footnote{\url{https://www.gov.uk/government/publications/standard-industrial-classification-of-economic-activities-sic}} of economic activities to the company on demand like `fund management activities' without the company getting verified and validated by the FCA.

While investigating the HYIPs manually, we found HYIPs that have photo-shopped the publicly available official certificate and made it look like they are registered as a company in the country. Fig~\ref{pdf:fake_certificate_visualhyip} is an example of an HYIP using a fake registration document of the UK Companies House, which we found on an HYIP web page.

This variable is a boolean and takes upon the value of true if we find a valid Companies House registration for the website.

\paragraph{Same UK Address}
There was a bit of overlap in addresses for these schemes, either from, likely, the same actor running multiple schemes or, less likely, different actors purportedly running schemes from the same address. We do not find evidence that these are default in software packages for HYIP schemes or kits, but this could also explain some overlaps. We also do not find duplicate addresses for schemes with non-UK addresses. This variable is true if the address is the exact same as another scheme's address.

\paragraph{Valid Goldcoders license} 
\href{https://www.goldcoders.com/}{\emph{Goldcoders}} is one of the most commonly used HYIP kit developers. This service provides the ability to make a customized HYIP website with little technical experience and includes all of the necessary functions for running an HYIP. We validate all the HYIP domains' Goldcoders license using the \href{https://www.goldcoders.com/?page=checkdomain}{\emph{Goldcoders}} website. We found that 286 out of our 450 HYIPs have valid licenses from this specific HYIP kit developer. This variable is true if the scheme has a valid Goldcoders license.

\paragraph{Payment Processors}
With the rise in digital currencies and wallets, investment frauds have found more than one way to accept payments from their investors. After investigating all the HYIPs, we found HYIPs accepting as high as fourteen different e-currencies and cryptocurrencies and the minimum being one. This shows that some HYIPs provide limited payment processors, in which case they generally choose the most common ones like Perfect Money or Bitcoin. On the other hand, some HYIPs provide multiple options assuming that it would attract more investors as the victim can invest in the currency he/she prefers. This variable has the number of currencies the HYIP accepts. 

We also have the most used processors - Perfect Money, Payeer, Bitcoin, Ethereum, and Litecoin as variables to analyze in-depth. If the scheme accepts the particular currency/payment processor, these variables are true.

\paragraph{Social Media Platforms}
HYIPs use these platforms to publicize themselves, seeking new investors and convincing them by virtually ``connecting" through social media. We found that 270 HYIPs have at least one social media platform used for advertising mentioned on their website. Some HYIPs were found to have more than one Telegram, Facebook or YouTube handles. This variable has the number of social media platforms the HYIP promotes itself on. 

We also have individual social media platforms - Telegram, Twitter, Facebook, Instagram, and Youtube as variables to analyze in-depth. These variables are true is the scheme uses a particular social media platform for advertising.

\paragraph{Contain Contact Number}
We collect all contact numbers provided by the HYIPs displayed on their website and add them to our dataset for analysis. We used the scraped HYIP website pages to collect the numbers automatically and found 95 HYIPs that provide a contact number. This variable is true if the scheme provides a contact number.

\paragraph{Lifetime (days)}
In order to find the lifetime of HYIPs in days, we used the scraped data from the aggregators \emph{\href{https://web.archive.org/web/20191231000251/https://hyip.com/forums/review/}{hyip.com}} and \emph{\href{https://web.archive.org/web/20210210124421/https://hyiprank.com/}{hyiprank.com}}.
We find the start date of the HYIP by checking the date mentioned in the threads by the aggregators. To find the last date that the HYIP was active on, we use the date of the last thread on the aggregators' website. As some HYIPs have been up and running, we use our last collection date, September 5, 2021, as their last ``seen alive" date.

While calculating lifetime for HYIPs using the start dates and end dates, 49 HYIPs were found to be still up and running as of September 5, 2021. For all of these still active HYIPs, we consider these data points to be ``censored," which is a statistical technique to deal with data where the exact number cannot be calculated. Here, because we do not know the precise end date, we consider this to be right-censored since the event is ``still alive."

\subsection{Ethical Considerations}

We were provided a full ethics approval from our University's Research Ethics Committee.

\paragraph{Reporting to Financial Conduct Authority (FCA)}
We reported all the up and running HYIP websites that claim to be registered in the UK to the FCA in line with the University Ethics Committee's recommendation. All regulated financial activities in the UK are required to be approved and verified by the FCA before conducting any regulated business in the UK. The FCA does not update the results of the investigations conducted by them. Furthermore, many of these schemes have ended naturally by the time of publication. Hence, we cannot provide any further results on the reported HYIPs.
\section{Analysis and Results}
\label{analysis}

This section investigates what factors affect the lifetime, and thus the expected revenue, of the high yield investment schemes. We focus on schemes purported to be registered in the United Kingdom.

\subsection{HYIP Lifetime overview}

\begin{figure}[!t]
    \centering
    \includegraphics[width=\linewidth]{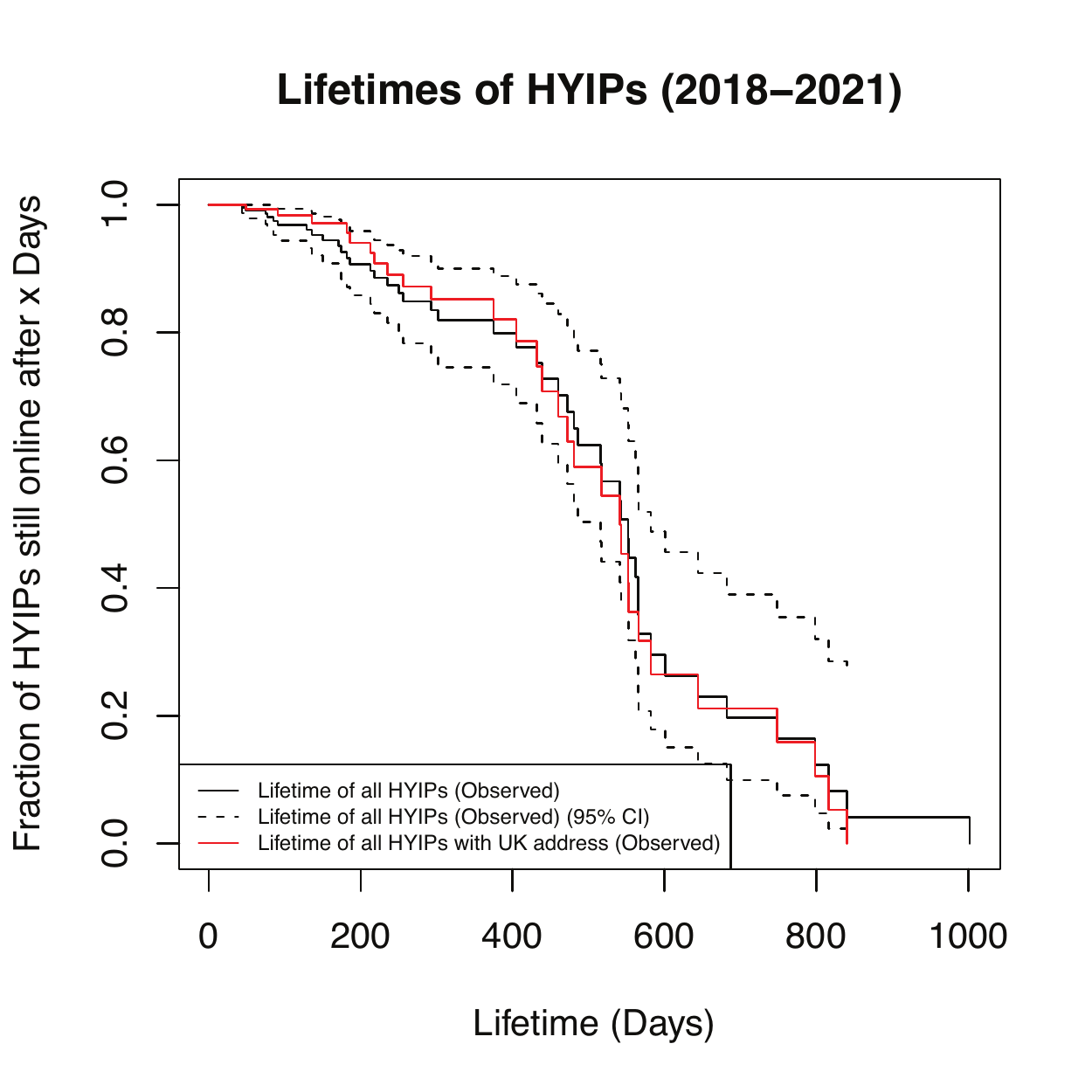}
    \caption{Survival Analysis of HYIP lifetimes ($N=450$) along with UK HYIPs ($N=283$); the chart shows that most HYIPs survive more than a year and UK HYIPs follow a similar trend ($p$ = 0.9 using a log rank test of difference).}
    \label{plot: survival_plot_withuk}
\end{figure}

We calculate the lifetime of all 450 HYIPs, where lifetime is an estimate of how long an HYIP has been online, detailed more in Section~\ref{sec:vars}. We use a method called survival analysis which allows us to analyze the lifetime of schemes, including schemes that were online at the time of data collection.

We had 49 schemes ``alive'' when our data collection stopped, and these 49 running HYIPs are `right-censored,' which indicates that they are still online at the end. We use a Kaplan-Meier estimator~\cite{kaplan1958nonparametric} to estimate the survival function $S(t)$ using the observed lifetime data. We use this to measure the fraction of HYIPs that collapse after a given date. Figure~\ref{plot: survival_plot_withuk} plots the lifetime of HYIPs in days with the overall survival probability. The dotted lines in the plot are the 95\% confidence interval. The red line shows the survival probability for HYIPs that claim to be from the UK.

The lifetime analysis by Moore et al. showed that the median lifetime of HYIPs is 28 days~\cite{moore:fc12}. They also found that one in four will last more than three months, and one in ten for more than ten months. In comparison, our research finds that the median lifetime of HYIPs is 43 days with a mean liftime of 118 days. 9.5\% of the HYIPs survive for more than a year. A possible reason for this difference might be that Moore et al. collected the data 11 years ago when HYIPs were comparatively new, and operators were scared to run it for a longer duration. We also use some collected data using the internet archive, which allows us to collect more data on more schemes but misses some short-run schemes.

Moore, Han and Clayton estimated that HYIPs attract at least \$6 million per month in revenue~\cite{moore:fc12}. Neisius and Clayton presented an in-depth model of the HYIP economy mentioning that the revenue, cost and profit of an average HYIP varies with the amount deposited~\cite{neisus:apwg}. As a rough cut, the length of time that a HYIP runs is proportional to the rate of return advertised and the the amount brought in. However, it is an estimate and future work in this field can better refine this relationship. Vasek and Moore found that the successful HYIP scams manage to pay out far less than they take in, and they do so consistently over time~\cite{vasek:fc15}. They found that longer running HYIPs earned significantly more money than shorter running HYIPs, though they noted an interaction between the type of HYIP and the length of the HYIP which might confuse some of these results.

\subsection{Explanatory Variables overview}

We study all variables mentioned in Table~\ref{tab:variables}. We individually use each social media platform and currency accepted for the social media platforms and currencies used variables. 

In order to see how these variables are correlated, we use a correlation matrix as shown in the appendix in Figure~\ref{tab:correlation}. The correlation matrix has been computed using Pearson and Spearman correlations. Positive correlations are displayed in blue, and negative correlations are in red. The color intensity and the size of the circle are directly proportional to the correlation coefficients. We note the correlation between different social media platforms and payment processors which leads us to consider them both together and separately.

\subsection{Analyzing Impact of UK address}
\label{sec:UKaddress}
\begin{figure}[!t]
    \centering
    \includegraphics[width=\linewidth]{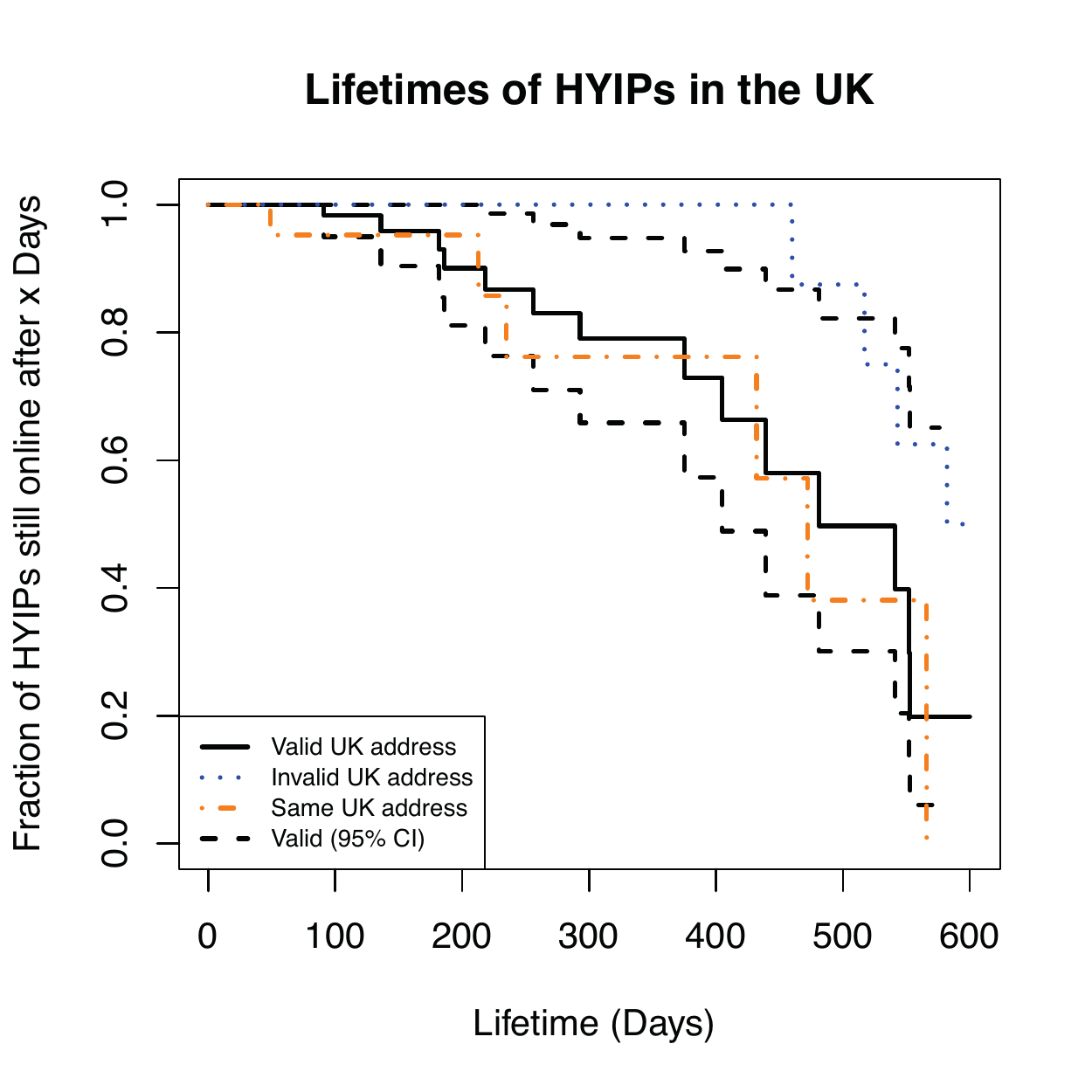}
    \caption{Survival Analysis of HYIP lifetimes for HYIPs operating in the UK that use a distinct address ($N=239$) where the lifetime differs based on whether the address is valid or not ($p-value = 0.01$ using a log rank test of difference).}
    \label{fig:ukaddsurv}
\end{figure}

Investigating HYIPs registered in the UK a bit further, we notice a difference in the behavior of Ponzi schemes that gave a valid or an invalid address (the process of detailing this distinction is outlined in Section~\ref{sec:vars}). There are many reasons why Ponzi scheme operators choose a particular address to advertise. We are not aware of any of these operators that have a presence in a particular address and have no reason to believe that these are a reflection of this, similarly to research looking at whether various cybercriminals use valid WHOIS contact information to register their websites~\cite{clayton:weis14,watters:ctcw13}. However, we hypothesize that different groups of people are registering their HYIPs under different protocols for choosing addresses and that this affects the lifetime of the various schemes. Figure~\ref{fig:ukaddsurv} shows that HYIPs with an invalid UK addresses are more likely to last past a year. The median survival is approximately 481 days for HYIPs with valid UK addresses and 613 days for HYIPs with invalid UK addresses, suggesting a good survival for HYIPs with invalid UK addresses. We observe no difference between addresses with valid UK addresses and those with duplicated UK addresses.

\subsection{Proportional Hazards Model}
\label{sec:ph}

In order to observe the varying effects of our variables on the lifetime of the HYIPs and more rigorously show the effects of using a valid or invalid UK address, we use a Cox proportional hazards model. Again, this uses a hazard function $h(t)$, which describes the probability of an event or its hazard $h$ if the subject survived up to that particular time point $t$~\cite{cox1972regression}. This is directly analogous to our survival analysis in the previous subsection. The hazard rate shows the probability that the HYIP will collapse, so an increased hazard rate means a greater risk of shutdown. In contrast, a decreased hazard rate means that the factor lowers its risk of immediate death.

\paragraph{Proportional Hazards: UK HYIPs} 

\begin{table*}[!t]
    \begin{center}
    \begin{tabular}{lcccl}
    \hline
    \multicolumn{1}{l}{Variable Name} & $coef$ & $exp(coef)$ & 95$\%$ $CI$ & \multicolumn{1}{c}{$p$ $value$}\\
    \hline
    Contain Contact Number & 0.3823 & 1.466 & (0.3089, 6.9555) & 0.6304 \\
    \textbf{Payeer} & \textbf{-1.543} & \textbf{0.2137} & (\textbf{0.0625, 0.7308}) & \textbf{0.0139} **\\
    Bitcoin & -0.7947 & 0.4517 & (0.0559, 3.6453) & 0.4557\\
    Ethereum & -1.157 & 0.3146 & (0.0488, 2.0289) & 0.2239\\
    Litecoin & 1.575 & 4.830 & (0.6604, 35.3217) & 0.1208\\
    Perfect Money & 18.34 & 92\,520\,000 & (0.000, Inf) & 0.9981 \\
    Telegram & -0.6892 & 0.5020 & (0.1719, 1.4657) & 0.2075 \\
    Twitter & -0.3678 & 0.6922 & (0.1603, 2.9888) & 0.6221 \\
    Facebook & 0.9300 & 2.535 & (0.4999, 12.8516) & 0.2615 \\
    \textbf{Instagram} & \textbf{2.080} & \textbf{8.002} & (\textbf{0.8496, 75.3615}) & \textbf{0.0691} *\\
    \textbf{Youtube} & \textbf{-1.829} & \textbf{0.1606} & (\textbf{0.0200, 1.2881}) & \textbf{0.0851} *\\
    Valid Goldcoders license & 0.1789 & 1.196 & (0.3086, 4.6341) & 0.7957 \\
    Registered in UK & 0.7996 & 2.225 & (0.3343, 14.8017) & 0.4083\\
    \textbf{Valid UK Address} & \textbf{1.729} & \textbf{5.636} & \textbf{(1.5183, 20.9194)} & \textbf{0.0098} ***\\
    Same UK Address & -0.1942 & 0.8235 & (0.1973, 3.4362) & 0.7899\\
    \hline
    \multicolumn{2}{l}{Concordance = $0.768$ $(se = 0.063)$} & \multicolumn{1}{c}{} & \multicolumn{1}{c}{} & \multicolumn{1}{c}{} \\
    \multicolumn{3}{l}{Likelihood ratio test = $23.98$ $on$ $15$ $df$, $p = 0.07$} & \multicolumn{1}{c}{} & \multicolumn{1}{c}{}\\
    \multicolumn{2}{l}{Wald test = $18.98$ $on$ $15$ $df$, $p = 0.2$} & \multicolumn{1}{c}{} & \multicolumn{1}{c}{} & \multicolumn{1}{c}{}\\
    \multicolumn{2}{l}{Logrank test = $22.3$ $on$ $15$ $df$, $p = 0.1$} & \multicolumn{1}{c}{} & \multicolumn{1}{c}{} & \multicolumn{1}{c}{}\\
    \hline
    \multicolumn{1}{c}{*** Significant at the $p=0.01$ level} & \multicolumn{1}{c}{}& \multicolumn{1}{c}{} & \multicolumn{1}{c}{}  & \multicolumn{1}{c}{}\\
    \multicolumn{1}{c}{** Significant at the $p=0.05$ level} & \multicolumn{1}{c}{}& \multicolumn{1}{c}{} & \multicolumn{1}{c}{}  & \multicolumn{1}{c}{}\\
    \multicolumn{1}{c}{* Significant at the $p=0.10$ level} & \multicolumn{1}{c}{}& \multicolumn{1}{c}{} & \multicolumn{1}{c}{}  & \multicolumn{1}{c}{}\\
    \hline
\end{tabular}
\caption{\label{tab:cox_hazard_uk}Cox proportional hazards model: Examining what affects the lifetime of HYIPs claiming to be from the UK ($N=283$); all variables.}
\end{center}
\end{table*}

\begin{table*}[!t]
    \begin{center}
    \begin{tabular}{lcccl}
    \hline
    \multicolumn{1}{l}{Variable Name} & $coef$ & $exp(coef)$ & 95$\%$ $CI$ & $p$ $value$\\
    \hline
    Contain Contact Number & 0.3284 & 1.3888 & (0.4228, 4.562) & 0.5883 \\
    \# of Payment Processors & 0.0679 & 1.0703 & (0.8711, 1.315) & 0.5177 \\
    \# of Social Media Platforms & -0.0782 & 0.9248 & (0.6472, 1.321) & 0.6677 \\
    Valid Goldcoders license & 0.6052 & 1.8317 & (0.6135, 5.469) & 0.2782 \\
    Registered in UK & 0.4299 & 1.5372 & (0.3475, 6.800) & 0.5709 \\
    \textbf{Valid UK Address} & \textbf{1.5219} & \textbf{4.5809} & (\textbf{1.3891, 15.106}) & \textbf{0.0124} **\\
    Same UK Address & 0.2347 & 1.2645 & (0.4005, 3.993) & 0.6891 \\
    \hline
    \multicolumn{2}{l}{Concordance = $0.694$ $(se = 0.06)$} & \multicolumn{1}{c}{} & \multicolumn{1}{c}{} & \multicolumn{1}{c}{} \\
    \multicolumn{3}{l}{Likelihood ratio test = $11.85$ $on$ $7$ $df$, $p = 0.1$} & \multicolumn{1}{c}{} & \multicolumn{1}{c}{}\\
    \multicolumn{2}{l}{Wald test = $9.06$ $on$ $7$ $df$, $p = 0.2$} & \multicolumn{1}{c}{} & \multicolumn{1}{c}{} & \multicolumn{1}{c}{}\\
    \multicolumn{2}{l}{Logrank test = $10.38$ $on$ $7$ $df$, $p = 0.2$} & \multicolumn{1}{c}{} & \multicolumn{1}{c}{} & \multicolumn{1}{c}{}\\
    \hline
    \multicolumn{1}{c}{** Significant at the $p=0.05$ level} & \multicolumn{1}{c}{}& \multicolumn{1}{c}{} & \multicolumn{1}{c}{}  & \multicolumn{1}{c}{}\\
    \multicolumn{1}{c}{* Significant at the $p=0.10$ level} & \multicolumn{1}{c}{}& \multicolumn{1}{c}{} & \multicolumn{1}{c}{}  & \multicolumn{1}{c}{}\\
    \hline
\end{tabular}
\caption{\label{tab:cox_hazard_uk_nonbi}Cox proportional hazards model: Examining what affects the lifetime of HYIPs claiming to be from the UK ($N=283$); measuring social media platforms and the number of payment processors collectively.}
\end{center}
\end{table*}

Table~\ref{tab:cox_hazard_uk} shows our regression results when the UK HYIP schemes are analyzed.
We observe that, as we have shown previously in Section~\ref{sec:UKaddress}, having a Valid UK address has a significant effect on the lifetime of an HYIP. Here we find that HYIPs with Valid UK addresses have an over 5 times higher hazard rate than those with invalid UK addresses, all else equal. This is statistically significant at the $p=0.01$ level. This means that having a valid UK address decreases the lifetime of the HYIPs and that HYIPs with an invalid UK address are more likely to last longer.

We observe other relationships that need further robustness checks to determine the impact of these findings.
We see that there is a survival advantage in advertising the use of Payeer. In fact, advertising the use of Payeer results in decrease of expected hazard rate by 78.63\% compared to those that do not advertise the use of Payeer, all else equal. This is statistically significant at the $p=0.05$ level. We hypothesize that this is because Payeer allows users to withdraw or deposit money into Payeer using cryptocurrency or other options (in multiple currencies). E-wallets are likely different than HYIPs that do not accept online payment processors and have different operating strategies.

We also find that using Instagram and Youtube to advertise the HYIPs to gain more investors have completely opposite significant effects. Promoting HYIPs on Instagram have an over 8 times higher hazard rate than those not using Instagram, all else equal. On the opposite side, HYIPs using Youtube decreases the hazard rate by 83.94\% compared to those that do not use Youtube, all else equal. These are statistically significant at the $p=0.1$ level. Our understanding is that investors do not use Instagram to find HYIPs and probably people report such accounts on Instagram. On the other hand, investors get to know detailed information through Youtube about the HYIPs functioning and usage, attracting more victims with a false sense of belief.

\paragraph{Proportional Hazards: additional models}
Given these results, we check for robustness by running additional regressions, both on our UK and our larger datasets. Table~\ref{tab:cox_hazard_uk_nonbi} looks into all UK variables along with other non-binary variables. We again find that having a valid UK address has a significant effect on HYIP's lifetime, adding to the robustness of this finding. HYIPs with Valid UK addresses have an over 4 times higher hazard rate than those with invalid UK addresses, all else equal. This is statistically significant at the $p=0.05$ level. This seconds the result about the valid UK address as seen previously in Table~\ref{tab:cox_hazard_uk} and also discussed in Section~\ref{sec:UKaddress}.

Cox regressions on the full dataset of UK and non-UK variables Table~\ref{tab:cox_hazard_bi}, \ref{tab:cox_hazard_bi_uk}, \ref{tab:cox_hazard_all_nonbi} show no significant effect of any of our other collected variables.

We were not able to gather data from the aggregator platforms after September 5, 2021 as their website were not alive since then. We expect these results to strengthen further after additional data collection on HYIPs over a longer period of time.

\subsection{Discussion: UK addresses do affect lifetime}

The survival analysis and Cox proportional hazards model from Section~\ref{sec:UKaddress} descriptively and Section~\ref{sec:ph} more rigorously confirm with the available data that purporting to have a valid address in the United Kingdom is a factor affecting the lifetime of the HYIPs. The collected data also shows that the three longest-running HYIPs have invalid UK addresses (excluding the one with longest lifetime that does not claim to be from UK).

This points to a strong likelihood that a Ponzi scheme website that purports to have a valid UK address to the UK Companies House has a higher hazard rate than those that do not.
We believe that this could be due to different attacker strategies. This is only one signal of additional investment in legitimacy indicators since coming up with a unique, valid UK address takes more code than simply falsifying one. This points to these sets of HYIP runners preferring schemes that likely offer higher rates of returns and die quicker.

There are several benefits of running these schemes in this manner, namely evading regulatory capture. Since any one scheme does not live long enough to capture enforcement interest, HYIP runners can play the system and start one scheme after another, with long enough time to let the advertising attracts enough customers, but not too long to collect law enforcement attention. This could also be a way to gather money from all available investors at the time since there is no evidence to support Ponzi scheme investors treating HYIPs more like a regular investment vehicle rather than a one-time gambling sport.
Collecting more data from other aggregators and looking into overall HYIP registrations outside the UK might provide new insights or strengthen the same result.
\section{Conclusion}
\label{conclusion}

High Yield Investment Programs (HYIPs) are a modern form of financial fraud as old as time~\cite{fbi_ic3}. We collected an updated data set of 450 new HYIPs from November 2020 to September 2021 with data as old as 2018 covering registration details of HYIPs, addresses, contact numbers, social media handles, and various currencies used by these HYIPs. We also verify that most HYIPs that claim to be in the United Kingdom are registered as a limited company through the UK Company Register known as Companies House.

We use Cox proportional hazards models to identify which variables have hazard rates with a statistically significant effect on the lifetime of the HYIPs.
There was a very strong support that having a valid UK address increases the hazard rate affecting the lifetime of the HYIPs.
We found some evidence towards the impact of social media. Promoting on Youtube helps prolong the HYIP lifetime where as Instagram decreases the HYIP's lifetime by over 8 times at some significant levels. We found limited evidence that for UK HYIPs, accepting Payeer decreases the risk of collapse by almost 84\%. 

\subsection{Recommendations}
\label{sec:rec}

We confirm in this paper that HYIPs register as a limited company in a few countries, including the United Kingdom and the United States of America. We provide the following recommendations:

\paragraph{Government organizations registering companies}
Currently, most organizations like the UK Companies House (UKCH) do not verify the documents submitted in order to register a company. We recommend verifying the documents that the applicants submit before registering a business as an official limited company. This should be easily possible and not need any new laws.
Organizations like UKCH should not provide a Standard Industrial Classification of economic activities (SIC) code for businesses submitting applications for finance-related activities. They should not be allowed to allocate codes like `Banks' or `Fund management activities' without the FCA's verification and approval. The current regulations need to be reiterated to implement this.

\paragraph{Government organizations regulating finance-related activities}
We recommend that agencies like the FCA, the U.S. Securities and Exchange Commission, and other similar organizations in their respective countries regulating finance-related activities, if not already, should collaborate with the organization that issues a company registration certificate. They should investigate businesses submitting applications to be registered as a limited company and thoroughly investigate applicants trying to register businesses conducting finance-related activities like investment companies or fund management. This might need a new regulation such that the stakeholders can work together to stop any suspicious companies from being registered officially.

\paragraph{Where to go from here}
While these HYIPs are currently registered in the UK, we are not fully convinced that the UK intervening here would cause these scams to collapse. Rather, HYIPs would purport to be from other countries or stop purporting to be from any country in particular. From a UK perspective, it is useful to have less of these scams associated with the country. But, from a global lens, UK-based regulations would likely reduce the credibility of these HYIPs, though, these would likely have limited effects. Hopefully, future work will be able to tease out the effects of UK-based regulation on the space. A more global approach might very well be necessary to stop the impact of these schemes.

\section*{Acknowledgment}
We would like to thank the reviewers for their helpful feedback on our work.
As part of the open-report model followed by the Workshop on Attackers \& CyberCrime Operations (WACCO), all the reviews for this paper are publicly available at \url{https://github.com/wacco-workshop/WACCO/tree/main/WACCO-2022}.

\bibliographystyle{IEEEtranN}
\bibliography{references}


\appendix

\begin{figure}[htbp]
    \centering
    \includegraphics[width=\linewidth]{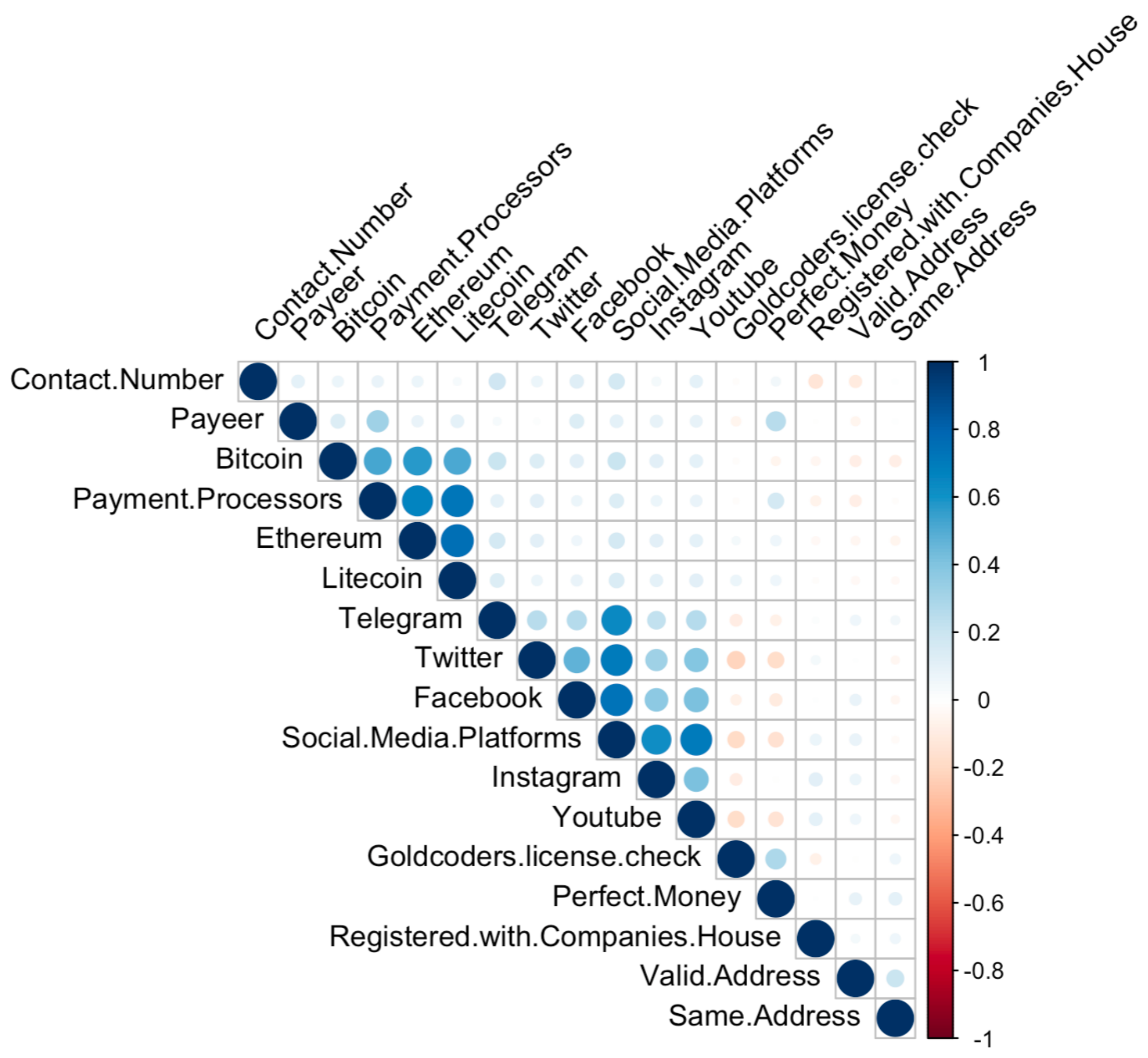}
    \caption{Correlation matrix plot of all variables (HYIPs $N=450$); the legend color shows the correlation coefficients and the corresponding colors.}
    \label{tab:correlation}
\end{figure}

\begin{table}[htbp]
    \begin{center}
    \begin{tabular}{lcc}
    \hline
    \multicolumn{1}{l}{Country}& \multicolumn{1}{c}{\# of HYIPs} & \multicolumn{1}{c}{\# of HYIPs} \\
    &\multicolumn{1}{c}{address listed} & \multicolumn{1}{c}{registered}\\
    \hline
    United Kingdom & 283 & 250\\
    Hong Kong & 7 & 7\\
    Republic of Seychelles & 1 & 1\\
    Saint Vincent and the Grenadines & 1 & 0 \\
    Republic of the Marshall Islands & 2 & 2\\
    United States of America & 9 & 4\\
    Australia & 9 & 3\\
    Germany & 3 & 0\\
    France & 1 & 1\\
    Netherlands & 1 & 0\\
    Belize & 3 & 2\\
    New Zealand & 2 & 1\\
    British Virgin Islands & 3 & 1\\
    Singapore & 2 & 0\\
    Belgium & 1 & 0\\
    Peru & 1 & 0\\
    Panama & 1 & 0\\
    Not Reported & 120 & \textbf{--}\\
    \hline
    \end{tabular}
    \caption{\label{tab:HYIP_distribution_countries}HYIP distribution in various countries ($N=450$); most HYIPs claim to be from the UK and are officially registered.}
    \end{center}
\end{table}

\begin{figure*}[htbp]
\centering
\includegraphics[width=\linewidth]{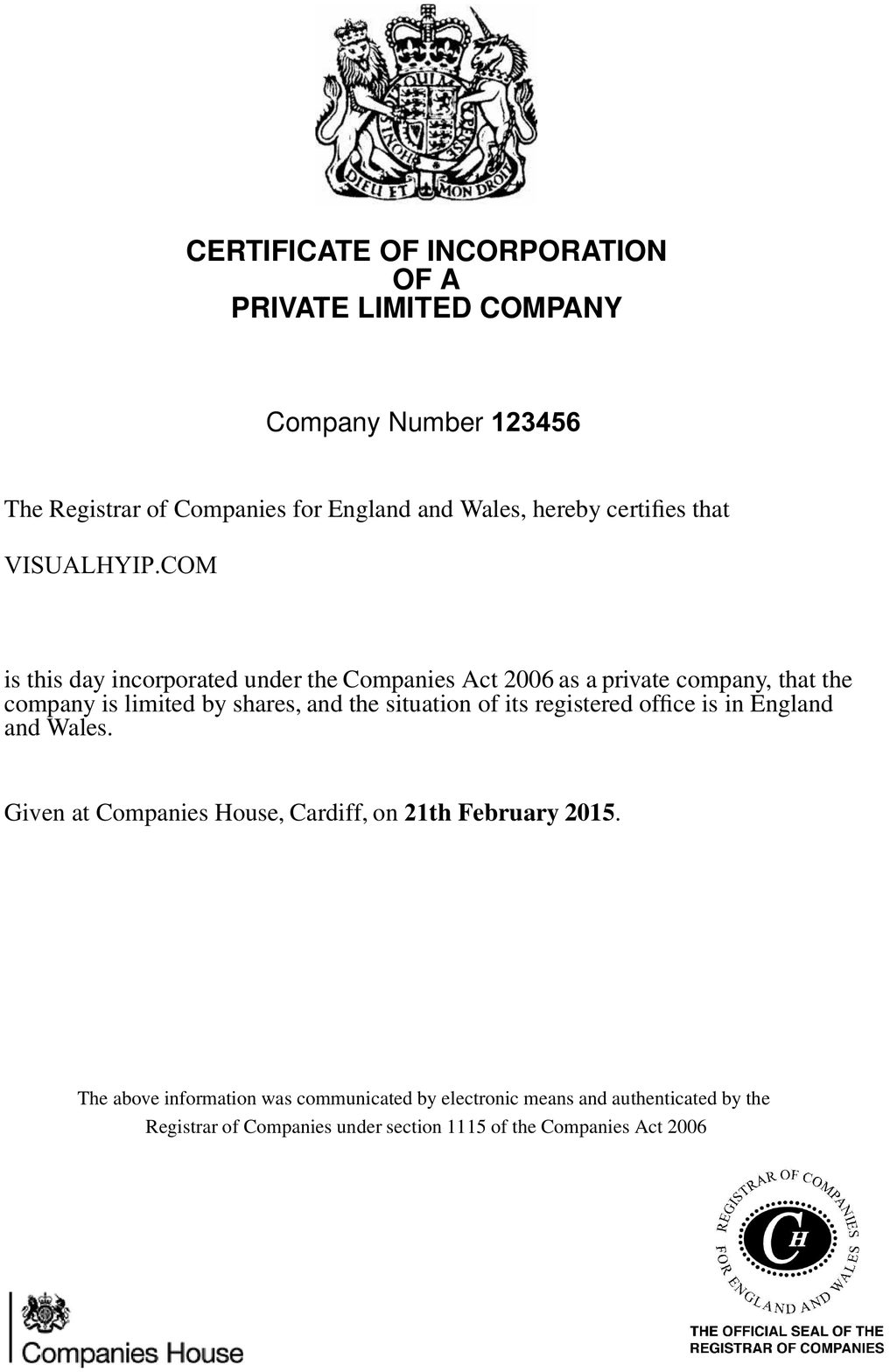}
\caption{Photo-shopped UK Companies House registration certificate with `Company Number 123456' provided by \url{https://visualhyip.com/}.}
\label{pdf:fake_certificate_visualhyip}
\end{figure*}

\begin{table*}[htbp]
    \begin{center}
    \begin{tabular}{lcccc}
    \hline
    \multicolumn{1}{l}{Variable Name} & $coef$ & $exp(coef)$ & 95$\%$ $CI$ & $p$ $value$\\
    \hline
    Contain Contact Number & 0.4450 & 1.5606 & (0.6576, 3.704) & 0.313\\
    Payeer & -0.4083 & 0.6648 & (0.2880, 1.535) & 0.339\\
    Bitcoin & -0.6028 & 0.5473 & (0.1740, 1.721) & 0.302\\
    Ethereum & -0.9438 & 0.3891 & (0.1092, 1.386) & 0.145\\
    Litecoin & 0.7852 & 2.1929 & (0.5916, 8.129) & 0.240\\
    Perfect Money & -0.6909 & 0.5011 & (0.1542, 1.629) & 0.251\\
    Telegram & -0.0785 & 0.9245 & (0.4531, 1.886) & 0.829\\
    Twitter & -0.0656 & 0.9365 & (0.3316, 2.645) & 0.901\\
    Facebook & 0.1568 & 1.1697 & (0.3930, 3.481) & 0.778\\
    Instagram & 0.6921 & 1.9979 & (0.4883, 8.174) & 0.336\\
    Youtube & -0.6119 & 0.5423 & (0.1710, 1.720) & 0.299\\
    Valid Goldcoders license & 0.1646 & 1.1790 & (0.5593, 2.485) & 0.665\\
    \hline
    \multicolumn{2}{l}{Concordance = $0.621$ $(se = 0.052)$} & \multicolumn{1}{c}{} & \multicolumn{1}{c}{} & \multicolumn{1}{c}{} \\
    \multicolumn{3}{l}{Likelihood ratio test = $11.05$ $on$ $12$ $df$, $p = 0.5$} & \multicolumn{1}{c}{} & \multicolumn{1}{c}{}\\
    \multicolumn{2}{l}{Wald test = $11.11$ $on$ $12$ $df$, $p = 0.5$} & \multicolumn{1}{c}{} & \multicolumn{1}{c}{} & \multicolumn{1}{c}{}\\
    \multicolumn{2}{l}{Logrank test = $11.75$ $on$ $12$ $df$, $p = 0.5$} & \multicolumn{1}{c}{} & \multicolumn{1}{c}{} & \multicolumn{1}{c}{}\\
    \hline
\end{tabular}
\caption{\label{tab:cox_hazard_bi}Cox proportional hazards model: Examining what affects the lifetime of HYIPs ($N=450$); all non-location variables.}
\end{center}
\end{table*}

\begin{table*}[htbp]
    \begin{center}
    \begin{tabular}{lcccl}
    \hline
    \multicolumn{1}{l}{Variable Name} & $coef$ & $exp(coef)$ & 95$\%$ $CI$ & $p$ $value$\\
    \hline
    Contain Contact Number & 0.4234 & 1.5272 & (0.6377, 3.657) & 0.342\\
    Payeer & -0.4216 & 0.6560 & (0.2830, 1.521) & 0.326\\
    Bitcoin & -0.6479 & 0.5231 & (0.1628, 1.682) & 0.277\\
    Ethereum & -0.9763 & 0.3767 & (0.1049, 1.353) & 0.135\\
    Litecoin & 0.8017 & 2.2294 & (0.5995, 8.290) & 0.232\\
    Perfect Money & -0.7413 & 0.4765 & (0.1429, 1.589) & 0.228\\
    Telegram & -0.1352 & 0.8735 & (0.4074, 1.873) & 0.728\\
    Twitter & -0.0499 & 0.9513 & (0.3358, 2.695) & 0.925\\
    Facebook & 0.1447 & 1.1557 & (0.3854, 3.465) & 0.796\\
    Instagram & 0.7269 & 2.0688 & (0.4985, 8.586) & 0.317\\
    Youtube & -0.6232 & 0.5362 & (0.1676, 1.715) & 0.294\\
    Valid Goldcoders license & 0.1220 & 1.1298 & (0.5200, 2.455) & 0.758\\
    UK Address & 0.1608 & 1.1744 & (0.5398, 2.555) & 0.685\\
    \hline
    \multicolumn{2}{l}{Concordance = $0.618$ $(se = 0.053)$} & \multicolumn{1}{c}{} & \multicolumn{1}{c}{} & \multicolumn{1}{c}{} \\
    \multicolumn{3}{l}{Likelihood ratio test = $11.22$ $on$ $13$ $df$, $p = 0.6$} & \multicolumn{1}{c}{} & \multicolumn{1}{c}{}\\
    \multicolumn{2}{l}{Wald test = $11.13$ $on$ $13$ $df$, $p = 0.6$} & \multicolumn{1}{c}{} & \multicolumn{1}{c}{} & \multicolumn{1}{c}{}\\
    \multicolumn{2}{l}{Logrank test = $11.83$ $on$ $13$ $df$, $p = 0.5$} & \multicolumn{1}{c}{} & \multicolumn{1}{c}{} & \multicolumn{1}{c}{}\\
    \hline
\end{tabular}
\caption{\label{tab:cox_hazard_bi_uk}Cox proportional hazards model: Examining what affects the lifetime of HYIPs ($N=450$); all variables along with UK Address variable which teases out any effects of purporting to be registered in the UK.}
\end{center}
\end{table*}

\begin{table*}[htbp]
    \begin{center}
    \begin{tabular}{lcccc}
    \hline
    \multicolumn{1}{l}{Variable Name} & $coef$ & $exp(coef)$ & 95$\%$ $CI$ & $p$ $value$\\
    \hline
    Contain Contact Number & 0.2377 & 1.2683 & (0.5792, 2.778) & 0.552 \\
    \# of Payment Processors & -0.0861 & 0.9175 & (0.8251, 1.020) & 0.112\\
    \# of Social Media Platforms & -0.0228 & 0.9774 & (0.7676, 1.245) & 0.853\\
    Valid Goldcoders license & 0.3352 & 1.3982 & (0.7253, 2.695) & 0.317\\
    \hline
    \multicolumn{2}{l}{Concordance = $0.569$ $(se = 0.059)$} & \multicolumn{1}{c}{} & \multicolumn{1}{c}{} & \multicolumn{1}{c}{} \\
    \multicolumn{3}{l}{Likelihood ratio test = $4.83$ $on$ $4$ $df$, $p = 0.3$} & \multicolumn{1}{c}{} & \multicolumn{1}{c}{}\\
    \multicolumn{2}{l}{Wald test = $4.67$ $on$ $4$ $df$, $p = 0.3$} & \multicolumn{1}{c}{} & \multicolumn{1}{c}{} & \multicolumn{1}{c}{}\\
    \multicolumn{2}{l}{Logrank test = $4.7$ $on$ $4$ $df$, $p = 0.3$} & \multicolumn{1}{c}{} & \multicolumn{1}{c}{} & \multicolumn{1}{c}{}\\
    \hline
\end{tabular}
\caption{\label{tab:cox_hazard_all_nonbi}Cox proportional hazards model: Examining what affects the lifetime of HYIPs ($N=450$); measuring social media platforms and the number of payment processors collectively.}
\end{center}
\end{table*}



\end{document}